\def\expec#1{\langle#1\rangle}

\def\etal{{\frenchspacing\it et al.}}
\def\ie{{\frenchspacing\it i.e.}}
\def\eg{{\frenchspacing\it e.g.}}
\def\etc{{\frenchspacing\it etc.}}

\def\rn{\bibitem[]{}}
\def\rf#1;#2;#3;#4 {\bibitem[]{}#1, {\it #2}, {\bf #3}, #4.}
\def\rg#1;#2;#3;#4;#5 {\bibitem[]{}#1, {\it #2}, {\bf #3}, #4 (``#5").}

\def\beq#1{\begin{equation}\label{#1}}
\def\eeq{\end{equation}}
\def\beqa#1{\begin{eqnarray}\label{#1}}
\def\eeqa{\end{eqnarray}}
\def\eq#1{equation~(\ref{#1})}

\def\eqnum#1{~(\ref{#1})}

\def\spose#1{\hbox to 0pt{#1\hss}}
\def\simlt{\mathrel{\spose{\lower 3pt\hbox{$\mathchar"218$}}
     \raise 2.0pt\hbox{$\mathchar"13C$}}}
\def\simgt{\mathrel{\spose{\lower 3pt\hbox{$\mathchar"218$}}
     \raise 2.0pt\hbox{$\mathchar"13E$}}}
\def\simpropto{\mathrel{\spose{\lower 3pt\hbox{$\mathchar"218$}}
     \raise 2.0pt\hbox{$\propto$}}}

\def\realpart{\hbox{Re}\,}

\def\Fh{\widehat{F}}
\def\nbar{{\bar n}}

\def\vk{{\bf k}}
\def\r{{\bf r}}

\def\k{{\bf k}}

\def\r{{\bf r}}
\def\x{{\bf x}}
\def\vx{{\bf x}}

\def\C{{\bf C}}

\def\I{{\bf I}}

\def\PP{{\bf\Pi}}
\def\W{{\bf W}}
\def\Pt{{\tilde P}}

\def\Z{{\bf Z}}
\def\Zt{\tilde{\bf Z}}

\def\vzero{{\bf 0}}
\def\psih{\widehat{\psi}}
\def\nnorm{\eta}
\def\nnormh{\widehat{\eta}}
\def\vnnorm{{\bf\eta}}

\def\Pt{{\tilde P}}

\def\delt{\delta_r}

\def\sshot{\sigma_s}
\def\ed{\end{document}}

\def\figone{
\makebox{
\smallskip
\noindent
\parbox[l]{3.7truecm}{\footnotesize
{\bf Figure 1.}\\
The exact expression for $N(k)$ is plotted together with the
approximation of Park {\it et al} for a Gaussian 
weight function $\psi(\r)\propto\exp[-(r/R)^2/2]$,
$R=100 h^{-1}$Mpc.
}
\hglue0.7truecm
\parbox[r]{5.5truecm}{
\epsfxsize=5.5truecm\rotate[r]{\epsfbox{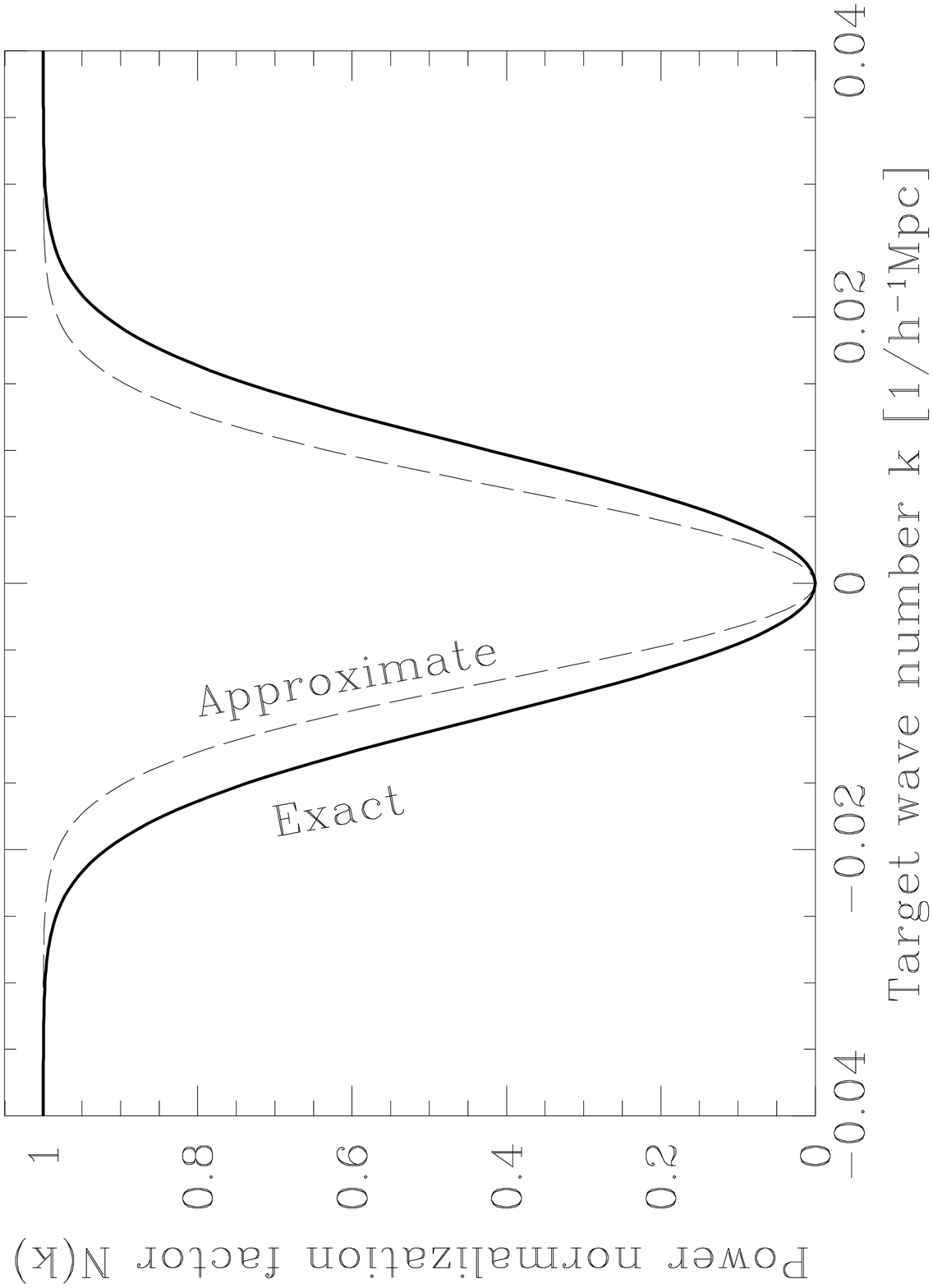}}
}
\smallskip
}
}

\documentstyle[editedvolume,namedreferences,epsf,rotate]{crckapb} 

\begin{opening}

\date{May 5, 1997}

\title{ANALYZING REDSHIFT SURVEYS TO MEASURE THE POWER SPECTRUM ON LARGE SCALES}

\author{MAX TEGMARK}

\institute{Institute for Advanced Study\\
           Princeton, NJ 08540; max@ias.edu}
           
\institute{Max-Planck-Institut f\"ur Physik\\  
F\"ohringer Ring 6, D-80805 M\"unchen}\\

\institute{Max-Planck-Institut f\"ur Astrophysik\\  
Karl-Schwarzschild-Str. 1, D-85740 Garching}\\

\end{opening}

\runningtitle{ANALYZING REDSHIFT SURVEYS}

\begin{document}


\begin{abstract}
Upcoming large redshift surveys potentially allow 
precision measurements of the galaxy power spectrum.
To accurately measure $P(k)$ on the largest scales, comparable
to the depth of the survey, it is crucial that finite volume effects
are accurately corrected for in the data analysis.
Here we derive analytic expressions for the one such effect that has not 
previously been worked out exactly: that of the so-called
integral constraint. 
We also show that for data analysis methods based on counts in cells,
multiple constraints can be included via simple matrix operations,
thereby rendering the results less sensitive to galactic
extinction and misestimates of the shape of the 
radial selection function.
\end{abstract}

\section{Introduction}

\vskip-11.97truecm
\hskip4.95truecm{\footnotesize 5}
\vskip11.50truecm

\vskip-18.00truecm
{$\>$}\\
{\it 
May 5, 1997.\\
To appear in {``Ringberg Workshop on Large-Scale Structure''},\\
ed. D. Hamilton (Kluwer, Amsterdam)
}
\vskip16.00truecm

\noindent
Observational data 
on galaxy clustering are 
rapidly increasing in both 
quantity and quality, which brings new challenges when it 
comes to data analysis.
As to quantity, redshifts had been published for a few thousand galaxies
15 years ago. Today the number is  
$\sim 10$\footnote[5]{Hubble Fellow}, 
and ongoing projects such as
the AAT 2dF Survey and the Sloan Digital Sky Survey (SDSS)
will raise it to $10^6$ within a few years.
Comprehensive reviews of past redshift surveys
are given by {\eg} Efstathiou (1994), 
Vogeley (1995), 
Strauss \& Willick (1995) and Strauss (1996), the last also including a
detailed description of 2dF and SDSS. 
As to quality, more accurate and uniform photometric selection criteria
(enabled by {\eg} the well-calibrated 5-band photometry of the SDSS)
reduce potential systematic errors.

This increased data quality makes it desirable to avoid 
approximations in the data analysis 
process and to use methods that can constrain cosmological 
quantities as accurately as possible, without bias. 
Here we will focus on how to correct for the finite volume of a
survey. As is well known, this causes the measured power spectrum to be
a convolution of the true power spectrum with some window function which
depends on the survey geometry and the data analysis method used. 
Exact expressions have been derived (see {\eg} Feldman, 
Kaiser \& Peacock 1994, hereafter ``FKP'') for the 
window function and its normalization for the case where the number density
of galaxies is assumed to be known {\it a priori}, but 
the more realistic case where the mean galaxy density is determined 
from the survey itself has thus far only been treated approximately
(Peacock \& Nicholson 1991; Park {\etal} 1994). 
The main purpose of this paper is to 
derive exact expressions for this important correction.

The methods for power spectrum estimation that have been 
proposed in the literature fall into two categories:
\begin{enumerate}
\item Direct Fourier methods
\item Pixelized methods
\end{enumerate}
The direct Fourier methods make use
of the exact position of each galaxy, whereas the other
methods start by ``pixelizing" the data set (by computing 
counts in cells or expansion coefficients for some 
set of functions), thereby reducing the problem to 
manipulating large vectors and matrices.
In Section~\ref{TraditionalSec}, we 
will derive the finite-volume correction for direct Fourier 
methods. The corresponding correction for pixelized methods
is given in Section~\ref{PixelizedSec}.

\section{Finite Volume Correction for Direct Fourier Methods}
\label{TraditionalSec}

\subsection{The power spectrum estimation problem}	 

It is customary (see {\eg} FKP) 
to model the observed galaxy 
distribution as a 3D Poisson process
$n(\r) = \sum_i \delta(\r-\r_i)$ with intensity 
$\lambda(\r) = \nbar(\r)[1+\delt(\r)]$.
The function $\nbar$ is the selection function of the galaxy
survey under consideration, {\ie}, $\nbar(\r)dV$
is the expected (not the observed) number of galaxies in 
a volume $dV$ about $\r$.
The density fluctuations $\delt$ are modeled as 
a homogeneous and isotropic 
(but not necessarily Gaussian) 
random field with power spectrum $P(k)$, 
and the power spectrum estimation 
problem is to estimate $P(k)$ given a realization of $n(\r)$.

\subsection{The direct Fourier approach}

Due to space limitations, the method summary below is very brief, and the 
interested reader is referred to FKP and Tegmark (1995, hereafter ``T95") 
for more detailed introductions to the various methods.

All direct Fourier methods not involving random numbers\footnote{
Including a random mock survey as in equation
(2.1.3) in FKP can never give minimal error bars, 
since inclusion of random numbers will always increase the 
variance of the estimator.
}
are specified by choosing a weight function $\psi(\r)$ in real 
space and a set of weights $w_i$ in Fourier space, as defined below.
They all involve the following two steps:
\begin{enumerate}
\item At a grid of points $\vk_i$ in Fourier space, 
fluctuation amplitudes are estimated by 
\beqa{FhDefEq}
\Fh(\vk_i) &\equiv& 
\int\left[{n(\r)\over\nbar(\r)}-1\right]\psi(\r)e^{-i\vk_i\cdot\r} d^3 r\nonumber\\
&=&\sum_j {\psi(\r_j)\over\nbar(\r_j)}e^{-i\vk_i\cdot\r_j}
- \psih(\vk_i).
\eeqa
(Here and throughout, hats denote Fourier transforms.)
\item The power $P$ at some given $k$-value, say $k_*$,
is estimated by squaring these fluctuation amplitudes, subtracting off
their shot noise bias, rescaling them to correct for the integral 
constraint, and averaging them with some weights $w_i$
that add up to unity:
\beq{PtDefEq}
\Pt(k_*) \equiv 
\sum_i w_i\left[{|\Fh(\vk_i)|^2-\sshot^2(\vk_i)\over N(\vk_i)}\right].
\eeq
\end{enumerate}
As we will show in Section~\ref{TradDerivationSec}, 
the new and exact expressions for the shot noise
and integral constraint corrections 
(when $\nbar$ is normalized so that $\Fh(\vzero)=0$) are 
\beqa{sshotDefEq}
\sshot^2(\vk)&=&\left(1+\left|{\psih(\vk)\over\psih(\vzero)}\right|^2\right)c_s(0)
- 2\,\realpart\left\{{\psih(\vk)^*\over\psih(\vzero)^*} c_s(\vk)\right\},\\
\label{NormDefEq}
N(\vk)&=&\left(1+\left|{\psih(\vk)\over\psih(\vzero)}\right|^2\right)f(0)
- 2\,\realpart\left\{{\psih(\vk)^*\over\psih(\vzero)^*} f(\vk)\right\},
\eeqa
where the functions $c_s$ and $f$ are defined by
\beqa{csDefEq}
c_s(\vk)&\equiv&\int{\psi(\r)^2\over\nbar(\r)}e^{-i\vk\cdot\r}d^3r,\\
f(\vk)&\equiv&\int \psi(\r)^2 e^{-i\vk\cdot\r}d^3r.
\eeqa
If the survey is volume limited, then $\nbar$ is independent of $\r$,
$c_s(\k)=f(\k)/\nbar$, and 
$\sshot^2(\vk)/N(\vk)=1/\nbar$. 

\subsection{Weighting the galaxies}

Four different choices of the galaxy weighting function $\psi$ have appeared 
in the literature:
\beqa{psiChoiceEq}
\psi(\r)&=&\left\{{\hbox{1 inside survey volume}\>\>\atop\hbox{0 outside survey volume}}\right.\\
\label{APMeq}
\psi(\r)&=&\nbar(\r)\\
\label{FKPeq}
\psi(\r)&=&{\nbar(\r)\over 1+\nbar(\r)P}\\
\label{T95eq}
\psi(\r)&=&\hbox{eigenfunction of }\left[\nabla^2-{\gamma\over\nbar(\r)}\right].
\eeqa
The first choice, {\ie}, weighing all galaxies in a survey volume 
equally, was employed by {\eg} Fisher {\etal} (1993).
The second choice was used for {\eg} the APM survey
(Baugh \& Efstathiou 1994) --- since redshifts were not measured, 
the radial galaxy weighting by default became the selection function
(moreover, modes could of course only be computed in the 
directions perpendicular to the line of sight).
The third choice is that advocated by Feldman, Kaiser \& Peacock (1994,
hereafter FKP), where $P$ denotes an {\it a priori} 
guess as to the power
in the band under consideration, and minimizes the variance in the limit 
when $k^{-1}\ll$ the depth of the survey. 
The fourth choice corresponds to the method of T95, 
and gives the narrowest window function for a given variance
(the constant $\gamma$ determines the tradeoff).

\subsection{Weighting the Fourier modes}

As to the weights in Fourier space, $w_i$, a common choice 
({\eg} FKP) is to perform a straight average of all modes in a spherical
shell with its radius centered on $k_*$, although when the survey volume is
anisotropic, a weighted average
giving smaller error bars can 
generally be obtained by solving 
a quadratic programming problem (T95).

\subsection{Window functions}

The expectation value of a power estimate 
$\Pt$ that has been corrected for the shot noise bias and 
the integral constraint can always be written as 
\beq{WindowDefEq}
\expec{\Pt} = 
\int W(k) P(k) dk,
\eeq
where the function $W$, known as the 
{\it window function}, has the property that
\beq{WindowNormEq}
\int_0^\infty W(k)dk=1.
\eeq
We can therefore think of $\Pt(k_*)$ as measuring a weighted average
of the true power spectrum, with $W$ specifying the weights
(for most methods, but not all, 
these weights are strictly non-negative as well).
The window function for a general direct Fourier method
is derived in Section~\ref{TradDerivationSec}, and is found to be
\beq{GeneralWeq}
\W(k) \propto \sum_i w_i \int |\psih_i(\k)|^2 k^2 d\Omega_k,
\eeq
where $\psi_i$ is given by \eq{psiiDefEq} and
the angular $\k$-integral is carried out over a spherical
shell of radius $k$. In the limit where $k^{-1}\ll L$,
where $L$ is the smallest
survey dimension, the 3D window function simplifies to 
$\W(k) \simpropto\sum_i w_i \int |\psih(\k-\k_i)|^2 k^2 d\Omega_k$.

\subsection{Derivation of the integral constraint correction}
\label{TradDerivationSec}


If we knew the selection function $\nbar(\r)$ {\it a priori}, 
before counting the galaxies in our survey, 
we would be able to probe
the power on the largest scales. For modes of wavelength
much larger than the survey volume, this would essentially 
correspond to counting the difference between the observed 
and expected number of galaxies in
our sample. 
Of course, we do not know $\nbar$ {\it a priori}, so our most 
accurate way of normalizing the selection function is by using 
the galaxies in the survey itself. 
When $\nbar$ is normalized in this way, 
naive application of \eq{FhDefEq} will give the artifact  
$\Fh(\k)\to 0$ as $\k\to 0$ because fluctuations on the scale of the survey
are forced to zero by definition (Peacock \& Nicholson 1991).

Let us assume that we know the shape of the selection function 
but not its normalization. To reflect this, we write
\beq{etaDefEq}
\nbar(\r)=\nnorm\nbar_0(\r),
\eeq
where $\nbar_0$ is our guess as to the shape and $\nnorm$ 
is an unknown normalization constant. 
If we had used $\nbar_0$ in place of the correct $\nbar$
in \eq{FhDefEq}, we would in general not obtain the desired result
$\expec{\Fh(\k_i)}=0$ but rather 
$\expec{\Fh(\k_i)}=(\nnorm-1)\psih(\vk)$, which 
would enter \eq{PtDefEq} as a systematic positive power bias.
It is the need to eliminate this problem that forces us to impose
an integral constraint. 
Let $\nnormh$ denote our estimate of $\nnorm$.
We will choose $\nnormh$ so that this bias vanishes, {\ie}, so that the
integral constraint
\beq{ImplicitEtaDef}
\int\left[{n(\r)\over\nnormh\nbar_0(\r)}-1\right]\psi(\r) d^3r=0
\eeq
holds, or explicitly, 
\beq{ExplicitEtaDef}
\nnormh\equiv {1\over\psih(\vzero)} 
\int {n(\r)\over\nbar_0(\r)}\psi(\r) d^3r=
{1\over\psih(\vzero)}\sum_j{\psi(\r_j)\over\nbar_0(\r_j)}.
\eeq
This is an unbiased estimator of the density 
normalization, since 
$\expec{\nnormh}=\nnorm$, the true value.
Substituting $\nbar(\r)=\nnormh \nbar_0(\r)$ 
and \eq{ExplicitEtaDef}
into \eq{FhDefEq}, we obtain 
\beqa{Fh2DerivEq}
\Fh(\vk_i)
&=&
{1\over\nnormh}
\left[\int{n(\r)\over\nbar_0(\r)}e^{-i\vk_i\cdot\r}\psi(\r) d^3 r
-\psih(\k_i)\nnormh\right]\nonumber\\
&=&
{1\over\nnormh}
\left[\int{n(\r)\over\nbar_0(\r)}e^{-i\vk_i\cdot\r}\psi(\r) d^3 r
-{\psih(\k_i)\over\psih(\vzero)}\int{n(\r)\over\nbar_0(\r)}\psi(\r) d^3 r\right]
\nonumber\\
&=&
{\nnorm\over\nnormh}\int {n(\r)\over\nbar(\r)}\psi_i(\r) d^3 r
\approx\int {n(\r)\over\nbar(\r)}\psi_i(\r) d^3 r,
\eeqa
where the function $\psi_i$ is defined by
\beq{psiiDefEq}
\psi_i(\r) \equiv
\left[e^{-i\vk_i\cdot\r}-{\psih(\k_i)\over\psih(\vzero)}\right]\psi(\r).
\eeq
Since we will have $\nnormh\approx\nnorm$ with a relative accuracy
$\Delta\nnormh/\nnorm$ of order 
$1/\sqrt{N}$, where $N$ is the number of galaxies 
in the survey, we can to a good approximation
treat $\nnorm$ as a known constant from here on and take
$\nnorm/\nnormh=1$ on the last line of \eq{Fh2DerivEq}.
Since $\psih_i(\vzero)=0$, we now have $\expec{\Fh(\vk_i)}=0$,
so we see that we have succeeded in eliminating the above-mentioned
power bias.
The price for this is slightly more complicated equations.  
Let us now derive the expressions for the
shot noise correction and normalization given in 
Equations\eqnum{sshotDefEq} and\eqnum{NormDefEq}.

Since $\nnormh\approx\nnorm$, 
we substitute the last expression of 
\eq{Fh2DerivEq} into Equation (3) of T95,
treating $\nbar=\nnorm\nbar_0$ as a known function,
which gives
\beq{F2ExpecEq}
\expec{|\Fh(\k_i)|^2} = 
{1\over(2\pi)^3}\int|\psih_i(\k)|^2 P(k) d^3k +
\int{|\psi_i(\r)|^2\over \nbar(\r)} d^3 r,
\eeq
Comparing this with \eq{WindowDefEq}
and \eq{PtDefEq}, we identify the three-dimensional 
window function as 
\beq{WindowEq2}
W(\vk)\propto|\psih_i(\k)|^2
\eeq
and see that the shot noise correction is 
\beqa{shotCorrEq1}
\sigma_s^2(\k_i)&=&\int{|\psi_i(\r)|^2\over \nbar(\r)} d^3 r\nonumber\\
&=&
\int\left|e^{-i\vk_i\cdot\r}-{\psih(\k_i)\over\psih(\vzero)}\right|^2
{\psi(\r)^2\over\nbar(\r)}d^3r,
\eeqa
and expanding the square completes our derivation of \eq{sshotDefEq}.
Performing an angular integral of \eq{WindowEq2}
completes the proof of \eq{GeneralWeq}.
The normalization coefficient $N(\k_i)$ of 
\eq{PtDefEq} is determined by the requirement that the
window function integrate to unity, {\ie},
$N(\k_i) = \int|\psih_i(\k)|^2 d^3 k/(2\pi)^3$.
Using Parseval's theorem, we obtain
\beqa{NderivationEq2}
N(\k_i) &=&\int |\psi_i(\r)|^2 d^3 r\nonumber\\
&=&\int\left|e^{-i\vk_i\cdot\r}-{\psih(\k_i)\over\psih(\vzero)}\right|^2
\psi(\r)^2 d^3r,
\eeqa
and expanding the square as above completes our derivation of \eq{NormDefEq}.

\figone

\subsection{How important is this correction?}

Let us evaluate the integral constraint correction factor $N(\k)$ 
for a couple of simple examples. We first note that for the 
special case of \eq{psiChoiceEq}, we 
have $\psi(\r)^2\propto \psi(\r)$. Hence $f(\k)\propto \psih(\k)$,
and \eq{NormDefEq} reduces to 
\beq{ParkApproxEq}
N(\vk)=\left(1-\left|{\psih(\vk)\over\psih(\vzero)}\right|^2\right)f(0),
\eeq
which we recognize as the approximation of 
Park {\etal} (1994).
For volume-limited surveys, 
the prescriptions given by equations\eqnum{APMeq},\eqnum{FKPeq} 
and\eqnum{T95eq} all coincide, so we see that this approximation 
becomes exact for the volume-limited case with these galaxy weighting 
schemes.
For flux-limited surveys, on the other hand, 
these schemes all give a decreasing weight function $\psi$,
since $\nbar$ decreases with distance.
For the simple Gaussian case 
$\psi(\r)=\exp[-(r/R)^2/2]/\pi^{1/4}R^{1/2}$,
\eq{NormDefEq} gives
\beq{GaussianNeq}
N(\k) = 1 + e^{-(Rk)^2} - 2e^{-{3\over 4}(Rk)^2},
\eeq
whereas the approximation\eqnum{ParkApproxEq} gives 
\beq{ParkNeq}
N(\k) = 1 - e^{-(Rk)^2}.
\eeq
A Taylor expansion shows that for $kR\ll 1$, 
the latter overestimates $N$ by a factor of two, 
as illustrated in Figure 1.

\section{Finite Volume Correction for Pixelized Methods}
\label{PixelizedSec}

\subsection{Pixelized methods}

Pixelized data analysis methods 
start by reducing the galaxy survey problem to one similar to that
occurring in cosmic microwave background (CMB) experiments:
estimating a power spectrum given noisy fluctuation measurements
in a number of discrete ``pixels". 
After this, the remaining steps are 
quite analogous to the CMB case, and involve mere linear algebra
(operations such as matrix inversion, diagonalization, {\etc}).
Let us define the overdensity in $N$ ``pixels" $x_1,...,x_N$ by 
\beq{xDefEq}
x_i \equiv\int\left[{n(\r)\over\nbar(\r)}-1\right]\psi_i(\r) d^3 r
\eeq
for some set of functions $\psi_i$.
Although the specific choices of
$\psi_i$ are irrelevant for our present discussion, 
common choices are to 
either make these functions fairly localized in real space
(in which case the pixelization is a generalized form of 
counts in cells) or fairly localized in Fourier space 
(in which case one refers to 
the functions $\psi_i$ as ``modes" and to $x_i$ as 
expansion coefficients).
Let us group the pixels $x_i$ into an $N$-dimensional vector $\vx$.
All proposed pixelized methods assume that
the mean and the covariance matrix of this pixel vector are
\beqa{xExpecEq}
\expec{\vx}&=&\vzero,\\
\expec{\vx\vx^t}&=&\C,
\eeqa
where $\C$ depends in some known way on the power spectrum. 
Once the problem has been cast in this form, the power spectrum
can be estimated using standard machinery, with either a 
brute force likelihood analysis (as in {\eg} White \& Bunn),
a Karhunen-Lo\`eve eigenmode analysis
(Karhunen 1947, Vogeley \& Szalay 1996; Tegmark, Taylor \& Heavens 1997)
or a direct quadratic analysis 
(Hamilton 1997ab; Tegmark 1997).

\subsection{Derivation of the integral constraint correction}

For pixelized methods of power spectrum estimation, 
the procedure for dealing with the integral constraint is 
quite analogous to that for direct Fourier methods. 
However, as we will now show, it is much simpler to implement.
For counts in cells, for example,
one simply removes the mean from all rows and columns of the covariance 
matrix $\C$ before proceeding with the analysis.
Because of this simplicity, one can, at an almost negligible 
numerical cost, take a more ambitious approach and allow for 
more than one unknown parameter in the selection function.
For instance, one 
can impose the constraints that the radial fluctuation average
equals zero for a few hundred different angular bins, thereby
eliminating the sensitivity to galactic extinction variations
on this scale, as well as requiring that the angular 
fluctuation average vanish for a number of radial bins 
to be insensitive to errors in estimating the precise
shape of $\nbar$.

Let us parametrize the true selection function $\nbar$ as
\beq{ParametrizedNbarEq}
\nbar(\r)=\sum_{j=1}^M \nnorm_j\nbar_j(\r),
\eeq
where $\nbar_j$ are known functions and
the ``nuissance parameters'' $\nnorm_j$, which we group into
an $M$-dimensional vector $\vnnorm$, are 
{\it a priori} unknown. Let $\nbar_0$ denote some
{\it a priori} estimate of $\nbar$.
Defining the ``uncorrected" pixels as
\beq{xpDefEq}
x'_i \equiv\int{n(\r)\over\nbar_0(\r)}\psi_i(\r) d^3 r,
\eeq
we find that 
\beq{xpExpecEq}
\expec{\x'}=\Z\vnnorm,
\eeq
where the $N\times M$-dimensional matrix $\Z$ is defined 
by
\beq{ZdefEq}
\Z_{ij}\equiv\int{\nbar_j(\r)\over\nbar_0(\r)}\psi_i(\r) d^3 r.
\eeq
This means that in general, $\expec{\x'}\neq 0$,
so the uncorrected data set does not satisfy \eq{xExpecEq}.
Instead, its statistical properties depend on the unknown nuissance 
parameters $\vnnorm$.
However, we can easily construct a new ``corrected" data set whose 
mean is independent of $\vnnorm$. Let us define 
\beq{xDefEq2}
\x\equiv\PP\x',
\eeq
where 
\beq{PPdefEq}
\PP\equiv \I-\Zt\Zt^t,
\eeq   
and $\Zt$ is a matrix whose rows form an orthonormal
basis ($\Zt^t\Zt=\I$)
for the space spanned by the rows of $\Z$.\footnote{
Such a matrix $\Zt$ is readily constructed by orthonormalizing
the rows of $\Z$ with a Gram-Schmidt or Cholesky procedure
(as in {\eg} Tegmark \& Bunn 1995).}
$\PP$ is a symmetric ($\PP^t=\PP$) projection matrix ($\PP^2=\PP$)
projecting onto the
subspace orthogonal
to the columns of $\Z$, {\ie}, $\PP\Z=\vzero$.
Our corrected data set $\x$ satisfies 
\eq{xExpecEq}, since
$\expec{\x}=\PP\Z\vnnorm =\vzero$.
Letting $\C'$ denote the covariance matrix of the uncorrected
data set, the corrected data will have the covariance matrix
\beq{xCovarEq2}
\C\equiv\expec{\x\x^t}=\PP\C'\PP.
\eeq
Once $\x$ and $\C$ have been computed, the rest of the pixelized 
analysis proceeds just as if there had been
no integral constraints.
The only complication is that $\C$ is now singular, 
having rank $N-M$ instead of $N$.
As shown in the Appendix of T97, the correct way to deal with
this is to replace all occurrences of $\C^{-1}$
(which is of course undefined) by the ``pseudo-inverse"
of $\C$, defined as 
\beq{PseudoInverseEq}
\PP\left[\C+\gamma\Z\Z^t\right]^{-1}\PP
\eeq
for some constant $\gamma\neq 0$. The result is independent of 
$\gamma$, but a good choice for numerical stability
is $\gamma\sim c/N$, where $c$ is the order of magnitude
of a typical matrix element of $\C$.


\bigskip
The author wishes to thank Josh Frieman, Andrew Hamilton, Michael Strauss and
Michael Vogeley for helpful comments on the manuscript.

\end{document}